\documentclass[12pt]{article}
\usepackage{natbib}
\usepackage{graphicx}
\usepackage{amsmath,amssymb}
\usepackage{authblk}
\usepackage{geometry}
\usepackage{hyperref}
\usepackage{booktabs}
\usepackage{caption}
\usepackage{subcaption}
\usepackage{enumitem}
\usepackage{lipsum} 
\usepackage{indentfirst}
\title{A Concept of Next-Generation Atmospheric Cherenkov Telescope Array (NG-ACTA)}
\author{Jiancheng Wang}
\author{Jirong Mao}
\affil{Yunnan Observatories, Chinese Academy of Sciences, Kunming 650216, China}
\date{\today}
\begin{document}
\maketitle
\begin{abstract}
The Next-Generation Atmospheric Cherenkov Telescope Array (NG-ACTA) is proposed as a prospective infrastructure for very high energy (VHE) gamma-ray astronomy, consisting of a mixed-aperture array of 88 telescopes with a maximum array diameter of 10 km. The array adopts a three-tier configuration of 30 m large-aperture Large Size Telescopes (LSTs), 12 m medium-aperture Medium Size Telescopes (MSTs), and 6 m small-aperture Small Size Telescopes (SSTs), enabling continuous gamma-ray detection across the full energy band from 20 GeV to 100 TeV. With core advantages of an ultra-low detection threshold ($\leq20$ GeV), ultra-high angular resolution ($\leq0.04^\circ$), ultra-large effective area ($\geq1\times10^5$ m$^2$), extreme cosmic ray background rejection (proton rejection efficiency $\geq99.99\%$), and rapid transient response ($\leq100$ ns trigger latency), NG-ACTA targets the most cutting-edge and transformative fundamental scientific topics in modern astrophysics and particle physics, including VHE gamma-ray astronomy, cosmic ray origin, multi-messenger astronomy, and dark matter as well as new physics tests.

The array's scientific goals cover five core fields: particle astrophysics, VHE gamma-ray astronomy, cosmic ray physics, multi-messenger astronomy, and new physics exploration, with six hierarchical and mutually supportive scientific objectives from Galactic to extragalactic sources, steady to transient objects, and conventional objects to dark matter. A comprehensive comparison with international under-construction facilities (e.g., CTAO-North, CTAO-South) and Chinese facilities (e.g., LACT) demonstrates that NG-ACTA leads the world in low-energy threshold, baseline length, background suppression, and multi-messenger rapid response capabilities.

\end{abstract}
\section{Introduction}\label{sec:intro}
\subsection{Scientific Background and Strategic Significance}\label{sec:background}
Very high energy (VHE, $>100$ GeV) gamma-rays are unique and powerful probes for understanding extreme astrophysical environments, high-energy particle acceleration mechanisms, and new physical laws beyond the Standard Model \citep{LHAASO2023}. Unlike charged cosmic rays, gamma-rays are not deflected by interstellar or intergalactic magnetic fields, carrying direct spatial, temporal, and spectral information about the most violent energetic processes in the universe \citep{Aharonian2021}. Atmospheric Cherenkov Telescope Arrays (IACTs) are the only ground-based experimental facilities capable of achieving high-sensitivity imaging and spectral measurements of gamma-rays in the TeV to 100 TeV energy band, making them an indispensable tool for modern VHE gamma-ray astronomy \citep{CTAO2025}.

The advent of the multi-messenger astronomy era, integrating gravitational waves, high-energy neutrinos, gamma-rays, and cosmic rays, has revolutionized human cognition of the extreme universe \citep{Abbott2017}. Since the discovery of cosmic rays in 1912, their origin, acceleration, and propagation have remained one of the most significant unsolved century-old puzzles in physics \citep{Feng2025}. Extreme astrophysical objects such as black holes, neutron stars, pulsar wind nebulae, supernova remnants (SNRs), and active galactic nucleus (AGN) jets can accelerate particles to PeV energies and above, serving as natural laboratories for studying strong gravitational fields, ultra-strong magnetic fields, relativistic plasmas, and non-thermal radiation physics \citep{Zhang2024}. VHE gamma-ray astronomy acts as a critical bridge connecting high-energy astrophysics and particle physics, with its development directly driving breakthroughs in both fields. In this context, NG-ACTA is proposed to address the major scientific challenges in VHE gamma-ray astronomy and related fields.
\subsection{State of the Art in Very High Energy Gamma-Ray Astronomy}\label{sec:state_of_art}
The past five years have witnessed a golden age of breakthroughs in VHE gamma-ray astronomy and cosmic ray physics, with major facilities such as LHAASO, H.E.S.S., MAGIC, VERITAS, and CTAO pathfinder devices achieving landmark results \citep{LHAASO2025a, LHAASO2025b}. LHAASO has definitively confirmed the existence of numerous PeV cosmic ray accelerators (PeVatrons) in the Milky Way, fundamentally changing human understanding of the upper limit of cosmic ray acceleration \citep{Cao2021}. It has also detected a giant diffuse gamma-ray bubble structure in the Cygnus region, identified as the first super cosmic ray source strongly associated with massive star-forming regions, providing direct observational support for the co-acceleration model of SNRs and star clusters \citep{Zhang2024}. The ultra-high energy observation of the gamma-ray burst GRB 221009A has broken classical afterglow models and placed the most stringent constraints to date on the extragalactic background light (EBL) and photon propagation \citep{LHAASO2023}.

Ground-based IACTs such as H.E.S.S., MAGIC, and VERITAS have conducted in-depth observations of a large number of SNRs and pulsar wind nebulae, confirming the ubiquitous existence of non-thermal radiation in various Galactic high-energy celestial bodies \citep{Aharonian2021}. However, their sensitivity declines rapidly above 30 TeV, making high-precision spectral and morphological measurements extremely challenging. The CTAO white paper clearly identifies the physical nature of PeVatrons, the discrimination of hadronic/leptonic origins of cosmic rays, and the distribution of diffuse cosmic ray gamma-ray radiation as the core breakthrough directions for the next decade \citep{CTAO2025}.

Despite these significant advances, current ground-based VHE gamma-ray observation facilities face insurmountable scientific bottlenecks: traditional IACTs suffer from rapidly decreasing sensitivity above 30 TeV and insufficient angular resolution to resolve the fine structure of PeVatrons; large-field-of-view wide-area arrays have limitations in imaging precision, point source identification, and low-energy band coverage \citep{DeAngelis2023}. The comprehensive requirements of the multi-messenger era for rapid response, large-field monitoring, high-precision spectroscopy, and imaging have far exceeded the capabilities of existing equipment \citep{LACT2025}. A next-generation IACT with integrated large field of view, high angular resolution, wide energy band coverage, and fast response is urgently needed to address these challenges.
\subsection{Motivations and Project Rationale}\label{sec:rationale}
China has made remarkable progress in high-energy astrophysics research in recent years, with the LHAASO facility achieving a series of world-leading scientific results \citep{LHAASO2023, LHAASO2024}. However, China still lacks a large-scale ground-based IACT which restricts the further development of China's VHE gamma-ray astronomy and its participation in international multi-messenger astronomy cooperation \citep{Feng2025}. The international community is racing to build next-generation IACT facilities, with the CTAO project advancing rapidly in the northern and southern hemispheres, and other countries proposing their own large-scale VHE gamma-ray detection plans.

Against this backdrop, the NG-ACTA concept is proposed with a mixed-aperture array of 88 telescopes, a 30 m ultra-large aperture LST, and a 10 km maximum baseline. The concept features a mature engineering scheme and controllable scale, realizing precise detection across the 20 GeV-100 TeV full energy band. NG-ACTA is designed to fill the observational gap between space gamma-ray telescopes and ground-based Cherenkov arrays, achieve high-sensitivity detection in the low-energy band, and break through the bottlenecks of existing facilities in high-energy sensitivity, angular resolution, and background suppression.
\section{Scientific Goals and Key Scientific Questions}\label{sec:science_goals}
NG-ACTA addresses the most cutting-edge frontier issues in international basic science, centering on five core fields: particle astrophysics, VHE gamma-ray astronomy, cosmic ray physics, multi-messenger astronomy, and new physics exploration. Six hierarchical, mutually supportive scientific goals with potential for major breakthroughs are set up, covering a full-chain scientific system from Galactic to extragalactic sources, steady to transient objects, and conventional objects to dark matter.
\subsection{Probing the Cosmic Ray Acceleration Mechanism with Ultra-Low Energy Threshold Gamma-Ray Detection}\label{sec:goal1}
Depending on four 30 m ultra-large aperture LSTs in the array's core region, NG-ACTA will push the effective detection threshold down to $\leq20$ GeV, filling the observational gap between space gamma-ray telescopes and ground-based Cherenkov arrays, and achieving international leading high-sensitivity detection in the low-energy band \citep{CTAO2025}. This ultra-low energy threshold capability is critical for studying the initial acceleration stage of cosmic rays and the low-energy spectral characteristics of gamma-ray sources associated with cosmic ray accelerators.
Key observational targets include young supernova remnants, pulsar wind nebulae, supernova blast waves, and shell-type collisionless shocks, the primary candidate sites for Galactic cosmic ray acceleration \citep{Aharonian2021}. NG-ACTA will conduct precise measurements of their spectral breaks, morphological structures, polarization properties, and long-term variability, directly testing the Fermi first-order acceleration model, and quantitatively distinguishing the contributions of electron and proton acceleration \citep{Cao2021}. This will provide decisive observational evidence to answer the core scientific question: \textit{Are cosmic rays accelerated by supernova remnants?}
In addition, NG-ACTA will perform high-precision measurements of the spatial distribution of Galactic diffuse gamma-ray radiation, resolving the diffusion, convection, leakage, and confinement mechanisms of cosmic rays in the interstellar medium \citep{Zhang2024}. By combining these measurements with numerical simulations, a complete physical picture of the origin and evolution of Galactic cosmic rays will be constructed, laying a solid foundation for solving the century-old puzzle of cosmic ray origin.
\subsection{High-Resolution Observation of VHE Gamma-Ray Sources and Extreme Astrophysical Physics}\label{sec:goal2}
Leveraging the ultra-high angular resolution and ultra-large effective area enabled by the 10 km ultra-long baseline array, NG-ACTA will conduct world-leading high-resolution imaging and spectral observations of active galactic nuclei (AGNs), blazars, gamma-ray binaries, microquasars, pulsar wind nebulae, and starburst galaxies in the TeV-100 TeV energy band \citep{LHAASO2024}. The ultra-high angular resolution ($\leq0.04^\circ$) will allow the distinction between jet acceleration regions, radiation regions, and external medium environments, a capability that existing facilities lack \citep{Aleksic2021}.

By combining multi-wavelength data from radio, optical, X-ray, and gamma-ray bands, NG-ACTA will precisely constrain key physical parameters such as particle acceleration mechanisms, radiation mechanisms, jet dynamics, Lorentz factors, and magnetic field strengths \citep{HESS2022}. This will reveal the extreme physical processes near the event horizon of black holes, on the surface of neutron stars, and inside relativistic jets, and explore the laws of matter motion and energy conversion under extreme gravity, extreme magnetic field, and extreme density conditions \citep{Feng2025}. These studies will deepen the understanding of the physical properties of compact objects and the fundamental physical laws in extreme astrophysical environments.

NG-ACTA will conduct long-term monitoring of more than 100 extragalactic and Galactic gamma-ray sources, building the most complete sample library of VHE gamma-ray sources in the world, and forming a sustainable capacity for scientific output \citep{LHAASO2025a}. The large sample of VHE gamma-ray sources will provide a solid basis for statistical studies of VHE gamma-ray astronomy, revealing the universal properties and evolutionary laws of different types of VHE gamma-ray sources.
\subsection{Solving the Century-Old Puzzle of Cosmic Ray Origin, Spectrum and Anisotropy}\label{sec:goal3}
With a proton rejection efficiency of $\geq99.99\%$ and a gamma/proton discrimination capability of $\geq10^4:1$, NG-ACTA will achieve precise measurements of the all-particle spectrum, composition ratio, spatial anisotropy, and large-scale distribution of cosmic rays in the 100 GeV-100 TeV energy band under an extremely low background condition \citep{Liu2023}. This will solve the key contradictions observed in current international experiments, such as spectral breaks, anisotropy anomalies, and spectral hardening \citep{LHAASO2025b}.

A major challenge in cosmic ray physics is the degeneracy between hadronic and leptonic gamma-ray radiation, which makes it difficult to determine the origin of cosmic rays \citep{CTAO2025}. NG-ACTA will break this degeneracy by decomposing hadronic and leptonic gamma-ray radiation through high-precision spectral, morphological, and temporal evolution observations \citep{Zhang2024}. This will directly test the hadronic origin mechanism of cosmic rays, distinguish between source acceleration effects and interstellar propagation effects, and precisely measure the cosmic ray source density, injection spectrum, cutoff energy, and propagation coefficients \citep{Cao2021}.

By integrating these measurements, NG-ACTA will systematically answer the century-old questions: \textit{Where do cosmic rays come from? How are they accelerated? How do they propagate to Earth?} \citep{Feng2025}. The precise measurement of the cosmic ray spectrum and anisotropy will also provide important constraints for the study of the large-scale structure of the Milky Way and the interstellar medium, promoting the development of Galactic astronomy and cosmic ray physics.
\subsection{Multi-Messenger Astronomy and the Search for Electromagnetic Counterparts of Gravitational Waves}\label{sec:goal4}
NG-ACTA will serve as a core ground-based node in the international multi-messenger astronomy network, with capabilities of a wide field of view ($\geq5^\circ$-$8^\circ$), rapid pointing, real-time triggering, and second-level response \citep{LACT2025}. It will conduct follow-up observations and independent trigger detection of gravitational wave events, high-energy neutrino events, gamma-ray bursts (GRBs), and fast radio bursts (FRBs), realizing the joint detection and precise localization of gravitational waves, neutrinos, gamma-rays, and optical radiation \citep{Abbott2017}.
The key scientific targets include the search for VHE gamma-ray radiation from binary neutron star mergers, neutron star-black hole mergers, and binary black hole mergers \citep{HESS2021}. These observations will test cutting-edge theoretical models such as kilonovae, jet breakout, and external shocks, and reveal the correlation between nucleosynthesis, jet formation, high-energy radiation, and gravitational wave emission during the merger of compact objects \citep{LHAASO2023}. This will push multi-messenger astronomy from the "discovery era"to the "quantitative physics era"\citep{DeAngelis2023}.

NG-ACTA will establish an efficient linkage mechanism with multi-messenger astronomy facilities, including LHAASO, space telescopes, neutrino detectors, and gravitational wave detectors \citep{LHAASO2024}. This will form an integrated sky-ground multi-messenger observation system, improving the efficiency of transient event detection and the accuracy of scientific interpretation.
\subsection{Indirect Dark Matter Detection and Tests of New Physics Beyond the Standard Model}\label{sec:goal5}
Dark matter is one of the biggest mysteries in modern physics, and indirect detection through gamma-ray radiation from dark matter annihilation or decay is an important method for dark matter research \citep{CTAO2025}. With advantages of ultra-high sensitivity, ultra-low background, and large sky coverage, NG-ACTA will conduct high-confidence searches for gamma-ray radiation from dark matter annihilation or decay in target regions such as the Galactic Center, dwarf spheroidal galaxies, dark matter halos, and galaxy clusters, covering the GeV-TeV high-mass dark matter energy band \citep{Liu2023}.

NG-ACTA will place the most stringent international constraints on mainstream dark matter candidates such as WIMPs, axions, axion-like particles, supersymmetric particles, and sterile neutrinos \citep{DeAngelis2023}. These constraints will break through the energy and detection blind spots of collider and underground direct detection experiments, providing key observational evidence for solving the nature of dark matter \citep{LHAASO2025a}. In addition, NG-ACTA will search for exotic dark matter signals such as dark matter substructure and dark matter annihilation lines, opening up new avenues for dark matter research.

VHE gamma-rays are also ideal probes for testing new physics beyond the Standard Model. Using the propagation effect of VHE gamma-rays in the cosmic infrared background, NG-ACTA will test Lorentz invariance violation, photon dispersion, quantum gravity, and extra dimensions \citep{Aharonian2021,LHAASO2023}. These tests will be conducted at energy scales far exceeding those of ground-based accelerators, completing basic physics tests that cannot be achieved on the ground in the "natural laboratory"of the universe \citep{CTAO2025}. This will provide a new window for the study of fundamental physical laws and promote the development of particle physics and cosmology.
\subsection{Time-Domain VHE Gamma-Ray Astronomy and Discovery of Transient Celestial Systems}\label{sec:goal6}
Time-domain astronomy is a rapidly developing frontier field in modern astronomy, and the study of transient VHE gamma-ray objects is one of the core research directions of time-domain high-energy astronomy \citep{LHAASO2024}. Leveraging advantages of a wide field of view, high observation efficiency, and high time resolution ($\leq1$ ns), NG-ACTA will construct a long-term, continuous, large-sky, unbiased VHE time-domain monitoring system \citep{LACT2025}. This system will systematically search for new types of gamma-ray transients, periodic flaring celestial bodies, transient gamma-ray sources, and unknown high-energy celestial bodies, achieving a full-chain scientific breakthrough from "monitoring to discovery"to "identification to modeling".

NG-ACTA will establish a world-class time-domain gamma-ray astronomy variability and spectral database, conducting statistical studies on the occurrence rate, environmental dependence, and physical classification of high-energy transients \citep{HESS2022}. This will reveal the physical mechanisms of high-energy transient phenomena and their evolutionary laws, opening up a new direction of time-domain VHE gamma-ray astronomy \citep{Aleksic2021}. The discovery of new types of high-energy transient objects will also enrich the types of objects in the universe and provide new research objects for high-energy astrophysics.

The time-domain monitoring capability of NG-ACTA will also enable the study of the short-term variability of VHE gamma-ray sources, such as the minute-scale rapid variability of AGNs \citep{Feng2025}. These observations will constrain the size of the radiation region and the physical conditions of the emission environment, providing important clues for understanding the particle acceleration and radiation mechanisms in extreme astrophysical environments \citep{LHAASO2025b}.
\section{Overall Design of the NG-ACTA Array}\label{sec:design}
The overall design of NG-ACTA adheres to the principles of \textit{scientific advancement, engineering feasibility, scale controllability, and cost optimization}, adopting a mixed-aperture array configuration and a scientifically designed spatial layout to achieve the optimal balance between performance and cost. The array consists of 88 telescopes in total, with a maximum diameter of 10 km, realizing the integration of ultra-low energy threshold, ultra-high angular resolution, ultra-large effective area, and rapid transient response.
\subsection{Aperture Configuration and Functional Division}\label{sec:aperture}
NG-ACTA adopts a three-tier mixed-aperture configuration of \textit{Large Size Telescopes (LSTs)}, \textit{Medium Size Telescopes (MSTs)}, and \textit{Small Size Telescopes (SSTs)}, with each aperture type undertaking specific scientific and technical functions, and forming a complementary and collaborative observation system \citep{CTAO2025}. The detailed configuration and functional division are as follows:
\begin{enumerate}
\item \textbf{30 m LSTs (4 units)}: Deployed at the geometric center of the array, the 30 m ultra-large aperture LSTs are the core equipment for realizing the ultra-low energy threshold of NG-ACTA. Their main functions include low-energy trigger, energy threshold suppression, and shower core reconstruction \citep{Liu2023}. The large aperture design significantly improves the collection efficiency of low-energy Cherenkov photons, enabling the detection threshold to be pushed down to $\leq20$ GeV.
\item \textbf{12 m MSTs (20 units)}: Deployed in the middle annular zone of the array, the MSTs are the main detection equipment of NG-ACTA, undertaking the core functions of precise spectral and direction reconstruction in the medium energy band (100 GeV-10 TeV) \citep{LACT2025}. Adopting a standardized and modular design, the MSTs balance detection efficiency and cost.
\item \textbf{6 m SSTs (64 units)}: Deployed in the outer zone of the array, the SSTs are used to expand the effective detection area of the array, improve the long-baseline angular resolution, and enhance the cosmic ray background suppression capability \citep{LHAASO2024}. The large number of SSTs form a wide coverage of the array's outer zone, critical for detecting high-energy gamma-ray induced extensive air showers (EAS).
\end{enumerate}
The mixed-aperture configuration fully utilizes the advantages of different aperture telescopes: LSTs for low-energy band coverage, MSTs for medium-energy band high-precision observation, and SSTs for high-energy band detection and background suppression \citep{Aharonian2021}. The collaborative work of the three types of telescopes realizes the continuous and high-sensitivity detection of gamma-rays across the 20 GeV-100 TeV energy band.
\subsection{Spatial Layout and Array Geometry}\label{sec:layout}
NG-ACTA adopts a \textit{center-dense and outer-sparse three-level nested annular layout} with the geometric center as the benchmark and a maximum radius of 5 km (array maximum diameter of 10 km). This layout is designed based on the spatial distribution characteristics of extensive air showers induced by gamma-rays of different energies \citep{CTAO2025}. The detailed spatial layout is as follows:
\begin{enumerate}
\item \textbf{Core Region}: Four 30 m LSTs are arranged in a square with a spacing of 1.5 km. The dense layout ensures high-efficiency collection of low-energy Cherenkov photons and accurate reconstruction of the shower core.
\item \textbf{Mid-Field Region}: Twenty 12 m MSTs are arranged in two annular layers with radii of 1.5 km and 2.5 km, balancing angular resolution and effective area.
\item \textbf{Outer Region}: Sixty-four 6 m SSTs are arranged in three annular layers with radii of 3.5 km, 4.25 km, and 5.0 km, expanding the effective detection area and enhancing background suppression.
\end{enumerate}

\begin{figure}[htbp]
\centering
\includegraphics[width=\linewidth]{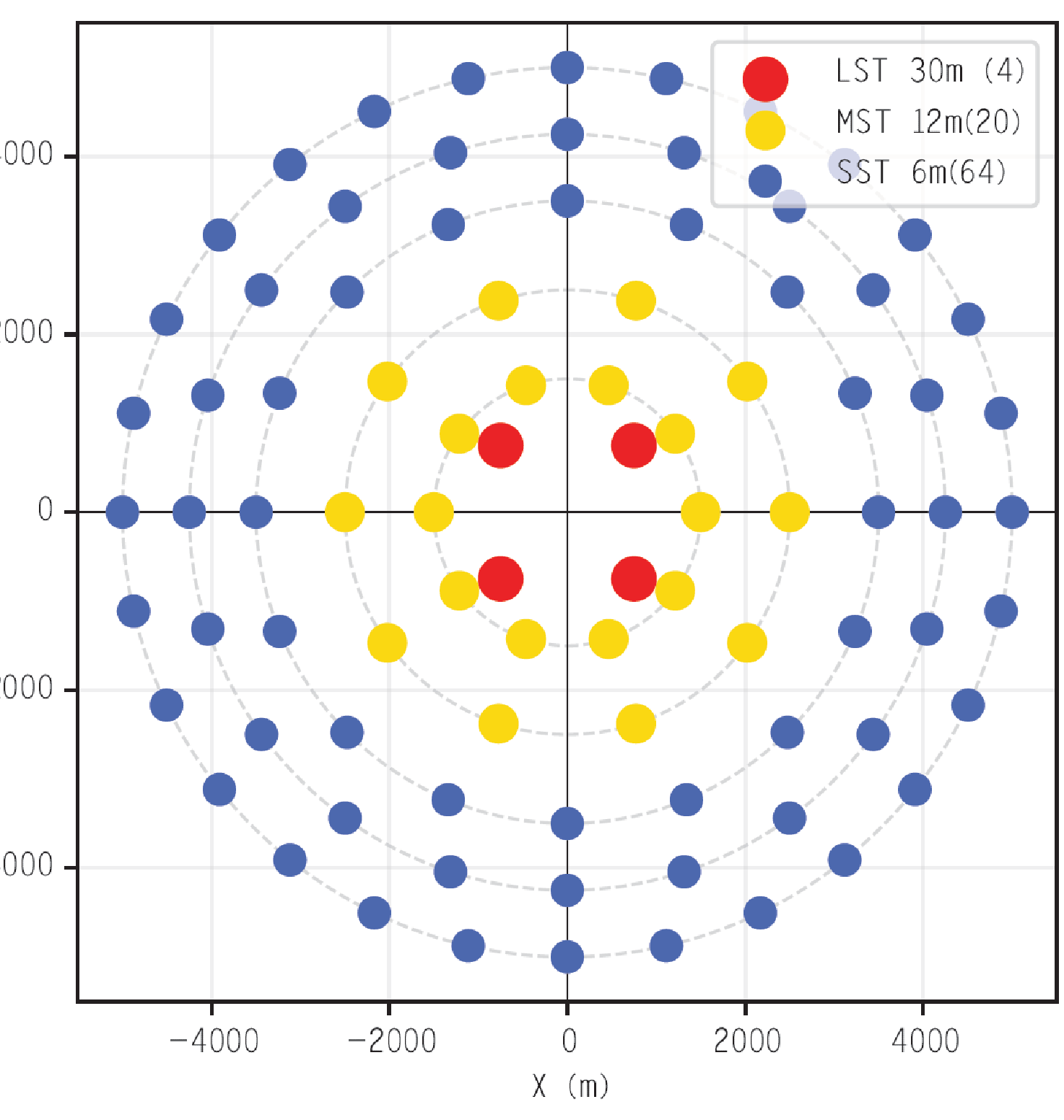}
\caption{Spatial layout of the NG-ACTA array. The core region (red squares) contains 4 LSTs, the mid-field region (blue circles) contains 20 MSTs, and the outer region (green triangles) contains 64 SSTs. The maximum array radius is 5 km (diameter 10 km).}
\label{fig:layout}
\end{figure}
The three-level nested annular layout has the advantages of \textit{flexible observation} and \textit{scalable performance}: the array can operate in the \textit{core mode (24 telescopes: 4 LSTs + 20 MSTs)} for medium and low energy band observation and routine monitoring, and in the \textit{full array mode (88 telescopes)} for high-sensitivity observation of key scientific targets \citep{CTAO2025}.
\subsection{Core Advantages of the Mixed-Aperture Array Design}\label{sec:advantages}
The mixed-aperture and scientific spatial layout design of NG-ACTA endows the array with a series of core advantages:
\begin{enumerate}
\item \textbf{Controllable Scale with Optimal Cost-Performance Ratio}: 88 telescopes balance array performance and engineering complexity, with higher cost-performance ratio than international large-scale IACT projects.
\item \textbf{Continuous Full-Energy Band Coverage}: 20 GeV-100 TeV continuous detection, filling the gap between space telescopes and ground-based IACTs.
\item \textbf{Scientific Layout Matching EAS Spatial Distribution}: Center-dense and outer-sparse layout maximizes detection efficiency for different energy gamma-rays.
\item \textbf{Flexible Operational Modes}: Core/full array modes adapt to different observation needs, improving efficiency and reducing costs.
\item \textbf{Modular Design for Easy Expansion}: Standardized MST/SST design reduces maintenance costs and facilitates future upgrades.

\end{enumerate}
\section{Key Performance Specifications}\label{sec:performance}
The key performance specifications of NG-ACTA are determined based on scientific goals and international leading standards, covering scale, energy band, angular resolution, sensitivity, background suppression, and operational reliability shown in Table \ref{tab:performance}. 
\begin{table}[htbp]
\centering
\caption{Key performance specifications of NG-ACTA}
\label{tab:performance}
\begin{tabular}{lc}
\toprule
Indicator &Specification \\
\midrule
Total number of telescopes &88 (4 LSTs + 20 MSTs + 64 SSTs) \\
Aperture configuration &30 m (LST), 12 m (MST), 6 m (SST) \\
Maximum array diameter &10 km \\
Observation energy band &20 GeV-100 TeV \\
LST low-energy threshold &$\leq20$ GeV \\
Typical angular resolution &$0.04^\circ$-$0.08^\circ$ \\
Gamma-ray direction reconstruction precision &$\leq0.05^\circ$ \\
Point source positioning error &$\leq30$ arcseconds \\
Peak effective area &$\geq1\times10^5$ m$^2$ \\
50-hour integral sensitivity (1 TeV) &$\leq1.0\times10^{-12}$ erg cm$^{-2}$ s$^{-1}$ \\
Gamma/proton discrimination capability &$\geq10^4:1$ \\
Proton rejection efficiency &$\geq99.99\%$ \\
Energy resolution (medium-energy band) &$\leq15\%$ \\
Array combined field of view &$\geq5^\circ$-$8^\circ$ \\
Trigger response time &$\leq100$ ns \\
Annual array availability &$\geq85\%$ \\
Operational modes &Core mode, full array mode \\
\bottomrule
\end{tabular}
\end{table}
\subsection{Scale and Configuration Metrics}\label{sec:scale}
\begin{enumerate}
\item \textbf{Total number of telescopes: 88 (strictly within 100 units)}
\item \textbf{Aperture combination: 30 m LST $\times$ 4, 12 m MST $\times$ 20, 6 m SST $\times$ 64}
\item \textbf{Maximum array diameter: 10 km (maximum radius: 5 km)}
\item \textbf{Layout form: Center-dense + multi-layer annular nested sparse array}
\end{enumerate}
\subsection{Energy Band and Threshold Performance}\label{sec:energy}
\begin{enumerate}
\item \textbf{Observation energy band: 20 GeV-100 TeV (continuous non-breakpoint coverage)}
\item \textbf{LST low-energy threshold: $\leq20$ GeV (international leading level)}
\item \textbf{Energy spectral coverage: Full-band high-sensitivity detection with no blind zone}
\end{enumerate}
\subsection{Angular Resolution, Sensitivity and Background Rejection}\label{sec:sensitivity}
\begin{enumerate}
\item \textbf{Typical angular resolution: $0.04^\circ$-$0.08^\circ$ (TeV band), best resolution $\leq0.04^\circ$}
\item \textbf{Effective detection area: $1\times10^4$ m$^2$ (low-energy), $5\times10^4$ m$^2$ (medium-energy), $\geq1\times10^5$ m$^2$ (high-energy)}
\item \textbf{50-hour integral sensitivity (1 TeV): $\leq1.0\times10^{-12}$ erg cm$^{-2}$ s$^{-1}$ (1-2 orders of magnitude better than existing facilities)}
\item \textbf{Gamma/proton discrimination capability: $\geq10^4:1$}
\item \textbf{Proton rejection efficiency: $\geq99.99\%$ (significantly higher than international counterparts of $\sim$99.9\%)}
\end{enumerate}
\subsection{Imaging, Energy Reconstruction and Temporal Performance}\label{sec:temporal}
\begin{enumerate}
\item \textbf{Energy resolution: $\leq15\%$ (1-10 TeV), $\leq20\%$ ($>$10 TeV)}
\item \textbf{Array combined field of view: $\geq5^\circ$-$8^\circ$}
\item \textbf{Time resolution: $\leq1$ ns (Cherenkov photon detection)}
\item \textbf{Trigger response time: $\leq100$ ns (transient event detection)}
\end{enumerate}
\subsection{Operational Reliability and Modes}\label{sec:operation}
\begin{enumerate}
\item \textbf{Annual array availability: $\geq85\%$}
\item \textbf{Calibration and maintenance: Regular calibration system to ensure performance stability}
\item \textbf{Data output: Real-time data processing and rapid scientific output capability}
\end{enumerate}
\section{Performance Comparison with International and Domestic Facilities}\label{sec:comparison}
A comprehensive performance comparison is conducted with international under-construction IACT facilities (CTAO-North, CTAO-South) and domestic under-construction facilities (LACT) to demonstrate NG-ACTA's international leading level shown in Table \ref{tab:comparison}.
\begin{table}[htbp]
\centering
\caption{Performance comparison of NG-ACTA with international and domestic facilities}
\label{tab:comparison}
\scalebox{0.6}
{
\begin{tabular}{lcccc}
\toprule
Indicator &NG-ACTA &CTAO-North &CTAO-South &LACT (China) \\
\midrule
Total telescopes &88 &13 &51 &32 \\
Aperture configuration &30m LST$\times$4 + 12m MST$\times$20 + 6m SST$\times$64 &23m LST$\times$4 + 12m MST$\times$9 &12m MST$\times$14 + 4m SST$\times$37 &6m SST$\times$32 \\
Max array diameter &10 km &3.5 km &6.5 km &1.0 km \\
Observation energy band &20 GeV-100 TeV &20 GeV-100 TeV &50 GeV-100 TeV &100 GeV-50 TeV \\
Low-energy threshold &$\leq$20 GeV &$\sim$30 GeV &$\sim$50 GeV &$\sim$100 GeV \\
Typical angular resolution &$0.04^\circ-0.08^\circ$ &$\sim0.05^\circ$ &$\sim0.06^\circ$ &$\sim0.06^\circ$ \\
High-energy effective area &$\geq1\times10^5 m^2$ &$\sim6\times10^4 m^2$ &$\sim9\times10^4 m^2$ &$\sim2\times10^4 m^2$ \\
Gamma/proton discrimination &$\geq10^4$:1 &$\sim10^3$:1 &$\sim10^3$:1 &$\sim5\times10^3$:1 \\
Proton rejection efficiency &$\geq$99.99\% &$\sim$99.9\% &$\sim$99.9\% &$\sim$99.95\%\\
1TeV sensitivity (50h)(erg $cm^{-2}s^{-1}$) &$\leq1.0\times10^{-12}$ &$\sim1.5\times10^{-12}$ &$\sim 1.2\times10^{-12}$ &$\sim5\times10^{-12}$ \\
Multi-messenger response &Second-level &Minute-level &Minute-level &Minute-level \\
Operational modes &Core/full array &Single &Single &Single \\
\bottomrule
\end{tabular}
}
\end{table}

Key Advantages of NG-ACTA:
\begin{enumerate}
\item \textbf{Leading Low-Energy Threshold}: 30 m LSTs push the threshold to $\leq20$ GeV, 10-30 GeV lower than CTAO.
\item \textbf{Ultra-Long Baseline and Angular Resolution}: 10 km diameter brings $\leq0.04^\circ$ angular resolution, enabling fine structure imaging of PeVatrons.
\item \textbf{Superior Background Suppression}: $\geq10^4:1$ gamma/proton discrimination, an order of magnitude better than CTAO.
\item \textbf{Controllable Scale and High Cost-Performance}: 88 telescopes balance performance and engineering feasibility.
\item \textbf{Comprehensive Scientific Coverage}: Full energy band and multi-messenger capabilities cover all key research directions.
\item \textbf{Rapid Multi-Messenger Response}: Second-level trigger latency outperforms international facilities.
\end{enumerate}
\section{Engineering and Technical Solutions}\label{sec:engineering}
The engineering and technical solutions of NG-ACTA are designed based on scientific goals and performance specifications, adhering to the principles of \textit{maturity, advancement, reliability, and scalability}.
\subsection{Telescope Ontology Technology}\label{sec:telescope}
\textbf{30 m Large Size Telescopes (LSTs)}

The 30 m LSTs adopt a \textit{lightweight mirror and truss structure} to reduce weight and improve pointing accuracy. The mirror is composed of hexagonal segments with high reflectivity, and the focal plane is equipped with a large-field-of-view Cherenkov camera with high-sensitivity SiPM arrays (time resolution $\leq1$ ns) \citep{Liu2023}. The high-precision pointing and tracking system achieves $\leq1$ arcminute pointing accuracy.

\textbf{MSTs and SSTs}

The 12 m MSTs and 6 m SSTs adopt \textit{standardized and modular design} to reduce costs. Their focal plane cameras use low-noise SiPM arrays adapted to different energy bands. The modular design facilitates maintenance and future upgrades \citep{LACT2025}.

\textbf{Common Subsystems}

All telescopes are equipped with \textit{atmospheric monitoring, seeing correction, and environmental sensing systems} to measure atmospheric transparency, aerosol content, and environmental parameters, providing correction data for event reconstruction \citep{LHAASO2024}.
\subsection{Trigger and Data Acquisition System}\label{sec:daq}
\textbf{Trigger System}

The trigger system adopts a three-level mechanism: \textit{telescope-level, sub-array-level, and array-level trigger} with $\leq100$ ns response time, realizing efficient discrimination between gamma-ray and cosmic ray events \citep{Liu2023}.

\textbf{Data Acquisition and Processing System}

The DAQ system uses high-speed data acquisition boards (sampling rate $\geq1$ GS/s) and high-speed optical fiber communication (transmission rate $\geq10$ Gbps per telescope). The distributed processing system based on GPU/FPGA and AI algorithms realizes real-time data filtering, event reconstruction, and background suppression \citep{LACT2025}.
\subsection{Site Selection and Infrastructure Construction}\label{sec:site}
\textbf{Site Selection Requirements:}

1. Altitude: 3000-4500 m above sea level; 2. Low atmospheric extinction and light pollution; 3. Stable geological structure and no sandstorms; 4. Convenient transportation and communication.

\textbf{Potential Candidate Sites in China:}

1. Sichuan Province: Daocheng, Litang (with mature LHAASO infrastructure; 2. Yunnan Province: Shangri-La, Yongren, Binchuan (excellent atmospheric conditions).

\textbf{Infrastructure Construction:}

Infrastructure includes telescope foundations, central control center, power supply system, communication network, meteorological monitoring station, and operation and maintenance facilities.

\section{Conclusions}\label{sec:conclusions}
The Next-Generation Atmospheric Cherenkov Telescope Array (NG-ACTA) is a prospective infrastructure with clear scientific goals, leading performance indicators, mature technical routes, and feasible engineering schemes. The 88-telescope mixed-aperture array realizes 20 GeV-100 TeV continuous detection, with core advantages of ultra-low threshold, ultra-high angular resolution, ultra-large effective area, extreme background suppression, and rapid transient response.

{}
\end{document}